\documentclass[3p,times,twocolumn]{elsarticle}

\usepackage{ecrc}

\volume{00}

\firstpage{1}

\journalname{Nuclear Physics B Proceedings Supplement}

\runauth{James P. Vary et al.}

\jid{nuphbp}

\jnltitlelogo{Nuclear Physics B Proceedings Supplement}

\usepackage{amssymb,amsmath}

\usepackage[figuresright]{rotating}
\usepackage{color}

\begin{document}

\begin{frontmatter}



\dochead{}

\title{Applications of Basis Light-Front Quantization to QED}


\author[label1]{James P. Vary}
\author[label1]{Xingbo Zhao}
\author[label2]{Anton Ilderton}
\author[label3]{Heli Honkanen}
\author[label1]{Pieter Maris}
\author[label4]{Stanley J. Brodsky}

\address[label1]{Department of Physics and Astronomy, Iowa State University, Ames, IA   50011 USA}
\address[label2]{Department of Applied Physics, Chalmers, SE-41296 G\"oteborg, Sweden}
\address[label3]{Dept. of Physics, Penn State University, University Park, PA 16802, USA}
\address[label4]{SLAC National Accelerator Laboratory, Stanford University, CA 94309, USA}

\begin{abstract}
Hamiltonian light-front quantum field theory provides a framework for calculating both static and dynamic
properties of strongly interacting relativistic systems.   Invariant masses, correlated parton amplitudes and time-dependent scattering amplitudes, possibly with strong external time-dependent fields, represent a few of the important applications. 
By choosing the light-front gauge
and adopting an orthonormal basis function representation, we obtain a large, sparse, Hamiltonian matrix 
eigenvalue problem for mass eigenstates 
that we solve by adapting ab initio no-core methods of nuclear many-body theory. In the continuum limit, the infinite matrix limit, 
we recover full covariance. Guided by the symmetries of light-front quantized theory, 
we adopt a two-dimensional harmonic oscillator basis for transverse modes that corresponds
with eigensolutions of the soft-wall anti-de Sitter/quantum chromodynamics (AdS/QCD) model obtained from
light-front holography. We outline our approach and present results for non-linear Compton scattering, evaluated non-perturbatively,  where a strong and time-dependent laser field accelerates the electron and produces states of higher invariant mass i.e. final states with photon emission.

\end{abstract}


\end{frontmatter}


\section{Introduction}
\label{Intro}

Recent interest in strong-field dynamics covers topics such as (a) the observed anomalous enhancement of lepton production in ultrarelativistic nuclear collisions at RHIC~\cite{Adare:2009qk}, (b) a prediction for photon yield depletion at the LHC~\cite{Tuchin:2010gx},   (c) effects of strong magnetic fields on Quantum ChromoDynamics (QCD)~\cite{Chernodub:2011mc,Bali:2011qj,Basar:2011by,Tuchin:2012mf}, and (d) studying both quantum electrodynamics (QED) and beyond-Standard-Model physics using next-generation lasers~\cite{DiPiazza:2011tq, Heinzl:2011ur, Jaeckel:2010ni}. This broad range of frontier applications points to the importance of developing new methods for solving quantum field theory (QFT) in nonperturbative and time-dependent domains.

For stationary state solutions alone, treating QFT in the nonperturbative regime remains a significant challenge.  There are immense challenges arising from the need to retain covariance and to develop numerical methods of sufficiently high precision, at practical computational cost.  A well-suited, flexible, and first-principles framework for this problem is provided by the Hamiltonian light-front formalism~\cite{Brodsky:1997de}, in which the theory is quantized on the light-front. The formalism is Lorentz frame independent and all observables are, in principle, accessible from the full time-dependent wavefunction.

Within the Hamiltonian light-front formalism, ``Basis Light-Front Quantization'' (BLFQ)~\cite{Vary:2009gt}, provides a nonperturbative calculational method for stationary states of QFT~\cite{Vary:2009gt,Honkanen:2010rc}. BLFQ provides the framework to diagonalize, after truncation, the full QFT Hamiltonian and yields the physical mass eigenstates and their eigenvectors. The eigenvectors may then be used to evaluate experimental observables. This approach therefore offers opportunities to address many outstanding puzzles in nuclear and particle physics~\cite{Brodsky:2011vc,Brodsky:2012rw}.

In this paper, we survey the logical extension of BLFQ, called {\it time-dependent} Basis Light-Front Quantization (tBLFQ), which provides the fully quantum, real time evolution of a chosen initial state under the influence of both quantum effects and applied background fields with arbitrary time-dependence. For details see~\cite{Zhao:2013cma,Zhao:2013jia}.

Although we will show a specific application here, the method is generally applicable to time evolution even in the absence of external fields where one is simply following the evolution of a chosen non-stationary state of the system. Here, we summarize the application of tBLFQ to ``strong field QED'', in which the external field models the high-intensity fields of modern laser systems. Such light sources now routinely reach intensities of $10^{22}$ W/cm$^2$ and offer prospects for investigating effects such as vacuum birefringence~\cite{Heinzl:2006xc}.


\section{Overview of Basis Light Front Quantization}
\label{Overview}

We define our 
light-front coordinates as $x^{\pm}=x^0 \pm x^3$, $x^{\perp}=(x^1,x^2)$,
where the variable $x^+$ is light-front time and $x^-$ is the longitudinal 
coordinate. We adopt the ``null plane" $x^+=0$ for our quantization surface.
Here we adopt 
basis states for each constituent that consist of transverse
2D harmonic oscillator (HO) states combined with discretized 
longitudinal plane waves. 
Successful anti-de Sitter-QCD models \cite{deTeramond:2008ht,Brodsky:2013ar} may also be viewed as
supporting this basis function approach \cite{Vary:2009gt,Vary:2009qz,Maris:2013qma,Vary:2013kma}.

The HO states are characterized by a radial quantum 
number $n$,  orbital quantum number $m$, and HO energy 
$ \Omega $. Working in momentum space, it is convenient to write the 
2D oscillator as a 
function of the dimensionless variable  
$\rho=\vert p^\perp\vert/\sqrt{M_0\Omega}$, and 
$M_0$ has units of mass.
The orthonormalized HO wave functions in polar coordinates 
$(\rho,\varphi)$ are then given in terms of the 
generalized Laguerre polynomials, $L_n^{|m|}(\rho^2)$, by
\begin{align}
\Phi_{nm}(\rho,\varphi)= 
\sqrt{\frac{2\pi}{M_0\Omega}}\sqrt{\frac{2n!}{(\vert m\vert+n)!}} 
e^{im\varphi}
\rho^{\vert m\vert} e^{-\rho^2/2}L^{\vert m\vert}_n(\rho^2),
\label{Eq:wfn2dHOchix}
\end{align}
with eigenvalues $E_{n,m}=(2n+|m|+1)\Omega$.
The HO wavefunctions have the
same analytic structure in both coordinate and momentum space, 
a feature they share in common with a plane-wave basis.

Our longitudinal modes, $\psi_{k}$, are defined for
$-L \le x^- \le L$ with periodic (antiperiodic) boundary conditions for the photon (electron), i.e.,
\begin{eqnarray}
  \psi_{k}(x^-) &=& \frac{1}{\sqrt{2L}} \, {\rm e}^{i\,\frac{\pi}{L}k\,x^-},
\label{Eq:longitudinal1}
\end{eqnarray}
where $k\in\{1,2,3,...\}$ ($k\in\{\frac{1}{2},\frac{3}{2},\frac{5}{2},...\}$). We neglect the photon zero mode. We take $L=2\pi$\,MeV$^{-1}$ throughout this paper. Then the full 3D single-particle basis state is
\begin{eqnarray}
  \Psi_{k,n,m}(x^-,\rho,\varphi) &=& \psi_{k}(x^-) \Phi_{n,m}(\rho,\varphi).
\label{Eq:totalspwfn}
\end{eqnarray}

Following Ref.\cite{Brodsky:1997de} we introduce the total invariant
 mass-squared $M^2$ for 
the low-lying physical states in terms of a Hamiltonian $H$ times a 
dimensionless number for the conserved total light-front momentum $K$ through
\begin{eqnarray}
M^2 =  P^+P^- - P_{\perp}P_{\perp} = KH - P_{\perp}P_{\perp}
\label{Mass-squared}
\end{eqnarray}
where the correction for total transverse center of mass kinetic energy
depends on the HO scale $1/\sqrt{M_0\Omega}$ and 
is easily removed from each solution.
\section{Electron Anomalous Magnetic Moment}
\label{AnomMo}
\begin{figure}[th!]
\centering
\includegraphics[width=0.52\textwidth]{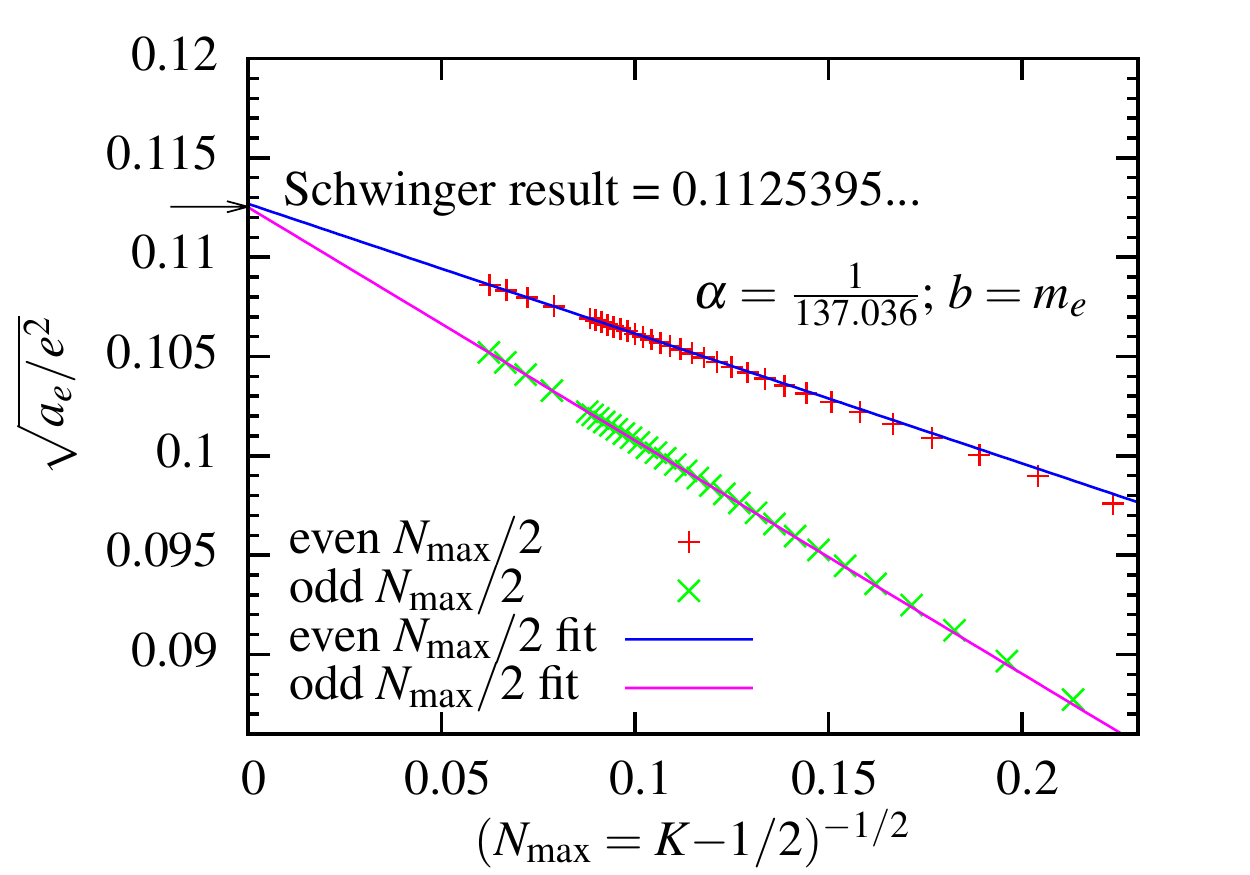}
\caption{(Color online) Square root of the electron anomalous magnetic moment ($a_e$), normalized by $e^2$, versus
the inverse square root of the HO basis cutoff $N_{\rm{max}}=K-\frac{1}{2}$.
Symbols are BLFQ results: squares (circles) for even (odd) $N_{\rm{max}}$/2, spanning $N_{\rm{max}}=18\ldots258$. Lines are linear extrapolations based on data points at $N_{\rm{max}}=K-\frac{1}{2}>100$ and agree favorably with the Schwinger result $\sqrt{{1}/{8\pi^2}} $~\cite{Schwinger:1948iu}.}
\label{anoma}
\end{figure}
 
We have solved for the electron's anomalous magnetic moment ($a_e$) when confined to an external trap and we have taken the limit where the trap vanishes~\cite{Honkanen:2010rc}.  Here we remove the trap altogether and follow procedures presented in more detail in Ref. \cite{Zhao:2014zzz}.  We adopt a Fock space of electron and electron-photon configurations consistent with a 
total angular momentum projection of $1/2$ and total HO quanta (sum over constituents' $2n+|m|$) 
limited by $N_{\rm{max}}$. 
The transverse infrared properties are still
regulated by the finite HO basis.  We also introduce a form of renormalization
that sets to unity the amplitude for the electron Fock space components of the 
interacting (physical) electron. With this prescription, as demonstrated in Fig. \ref{anoma}, the extrapolation to the infinite basis 
reproduces well the expected Schwinger result.

\section{BLFQ extended to include time-dependent processes}
We have recently extended the BLFQ to include the non-perturbative time-evolution of light-front systems~\cite{Zhao:2013cma,Zhao:2013jia} (tBLFQ). 
Light-front time evolution of quantum states ${|\psi;x^+\rangle}$ is governed by the Schr\"odinger equation,
\begin{equation}
\label{wave-eq}
	i\frac{\partial}{\partial x^+}{|\psi;x^+\rangle}= \frac{1}{2}P^-(x^+){|\psi;x^+\rangle} \;.
\end{equation}
\begin{figure}[!t]
\centering
\centering\includegraphics[width=0.45\textwidth]{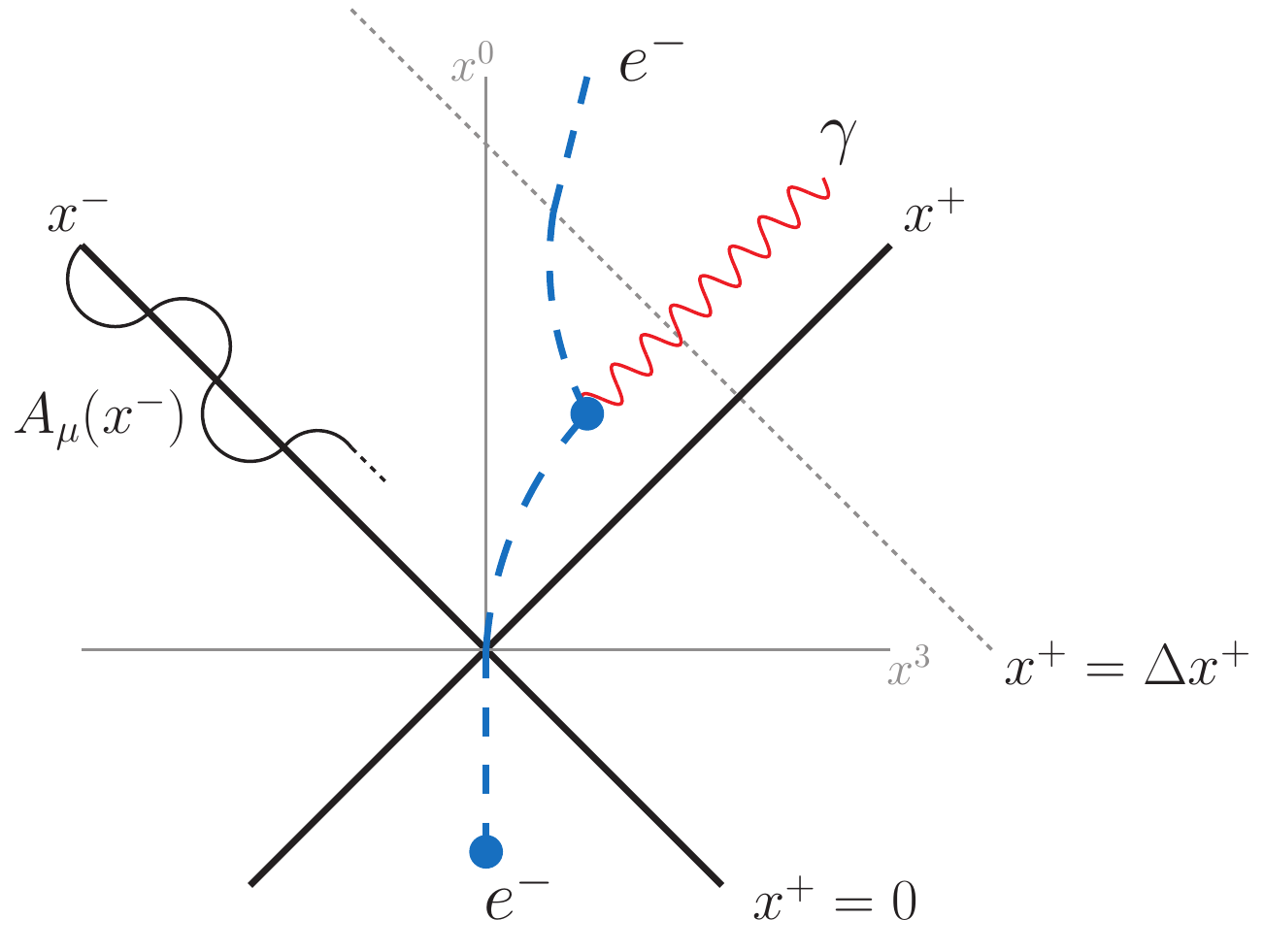}
\caption{An illustration of non-linear Compton scattering. An electron enters a laser field, is accelerated, and emits a photon. After emission the electron can be further accelerated until it leaves the field.}
\label{fig:nCs}
\end{figure}
In our QED demonstration case of non-linear Compton scattering illustrated in Fig. \ref{fig:nCs}, the Hamiltonian $P^-(x^+)$ contains two parts; $P^-_{QED}$ is the full Hamiltonian of QED, and $V(x^+)$ is the time-dependent interaction term introduced by the background field, so
\begin{equation}
	\label{H_interact}
	P^-(x^+)=P^-_{QED}+V(x^+) \;.
\end{equation}
It is natural to adopt the interaction picture where states are defined by
\begin{equation}
	{|\psi;x^+\rangle}_I = e^{\frac{i}{2}P^-_{QED}x^+}{|\psi;x^+\rangle} \;.
\end{equation}
These states obey the equation
\begin{equation}
\label{Schro-int}
	i\frac{\partial}{\partial x^+}{|\psi;x^+\rangle}_I= \frac{1}{2}V_I(x^+){|\psi;x^+\rangle}_I \;,
\end{equation}
in which $V_I$, `the interaction Hamiltonian in the interaction picture', is
\begin{equation}
	V_I(x^+) = e^{\frac{i}{2}P^-_{QED}x^+}V(x^+)e^{-\frac{i}{2}P^-_{QED}x^+} \;.
\end{equation}
The formal solution to (\ref{Schro-int}) is
\begin{equation}
	\label{i_evolve}
	{|\psi;x^+\rangle}_I  = \mathcal{T}_+ \exp\bigg(-\frac{i}{2} \int\limits_0^{x^+} V_I(x'^+)dx'^+\bigg){|\psi;0\rangle}_I .
\end{equation}

In tBLFQ, we solve Eq.~\ref{i_evolve} non-perturbatively. Some approximation must still be made in order to yield a (numerically) tractable system, but instead of resorting to perturbation theory, we use a Fock-space truncation (in the following example, we use the simplest nontrivial truncation, namely restricting to electron plus electron-photon states, along with 
$N_{\rm{max}}=8$).

\begin{figure*}[!t]
\centering
\includegraphics[width=0.49\textwidth]{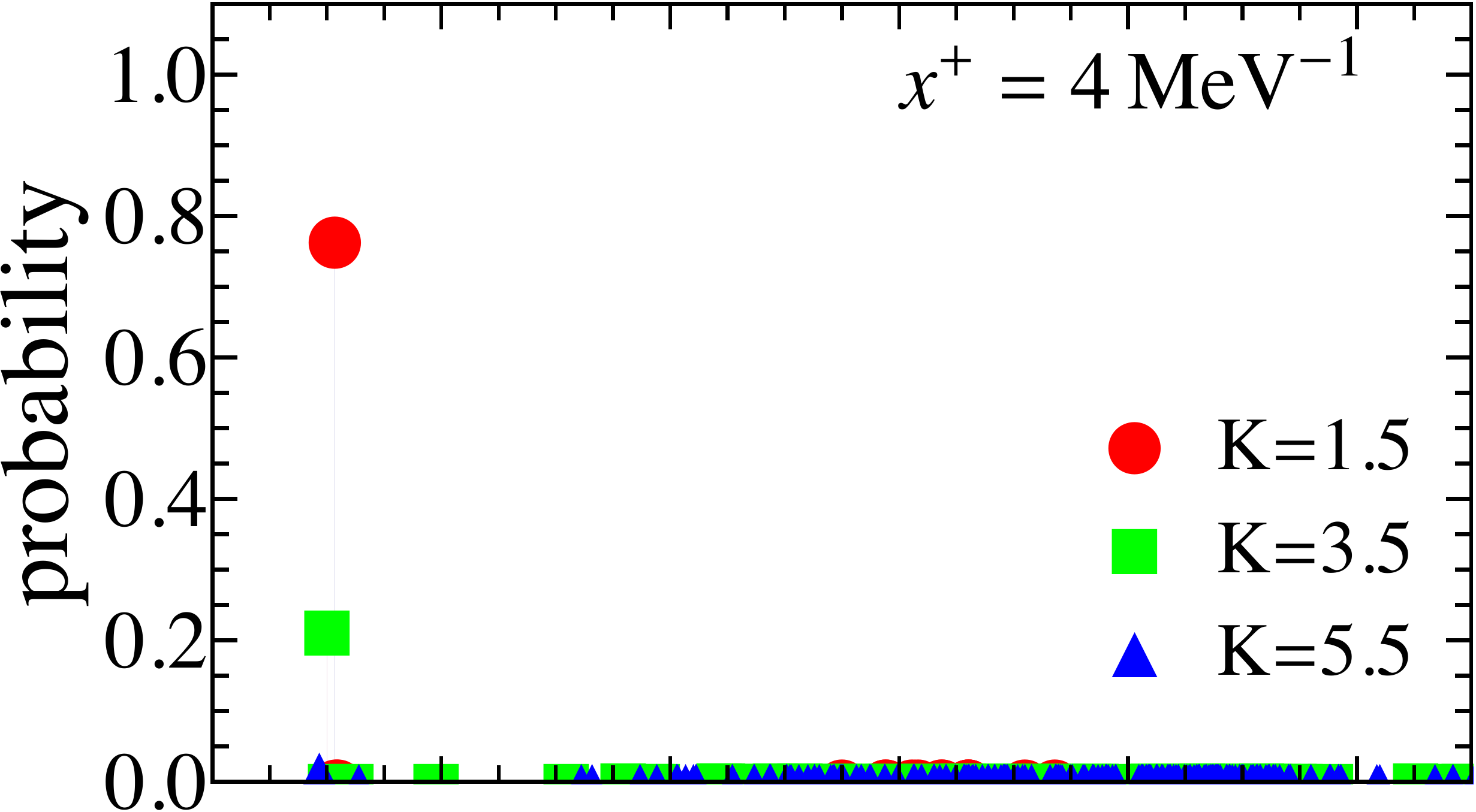}
\includegraphics[width=0.49\textwidth]{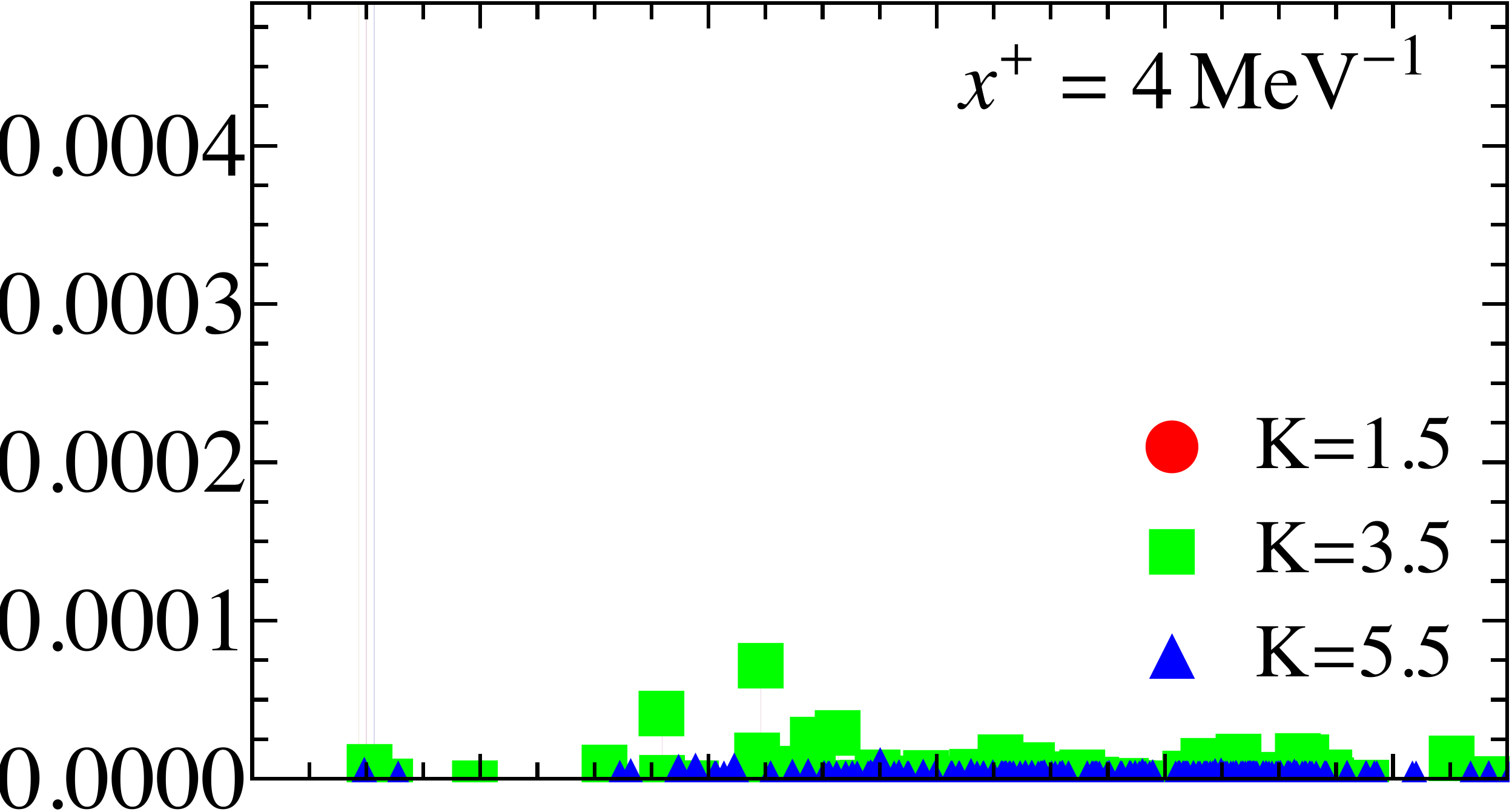}
\includegraphics[width=0.49\textwidth]{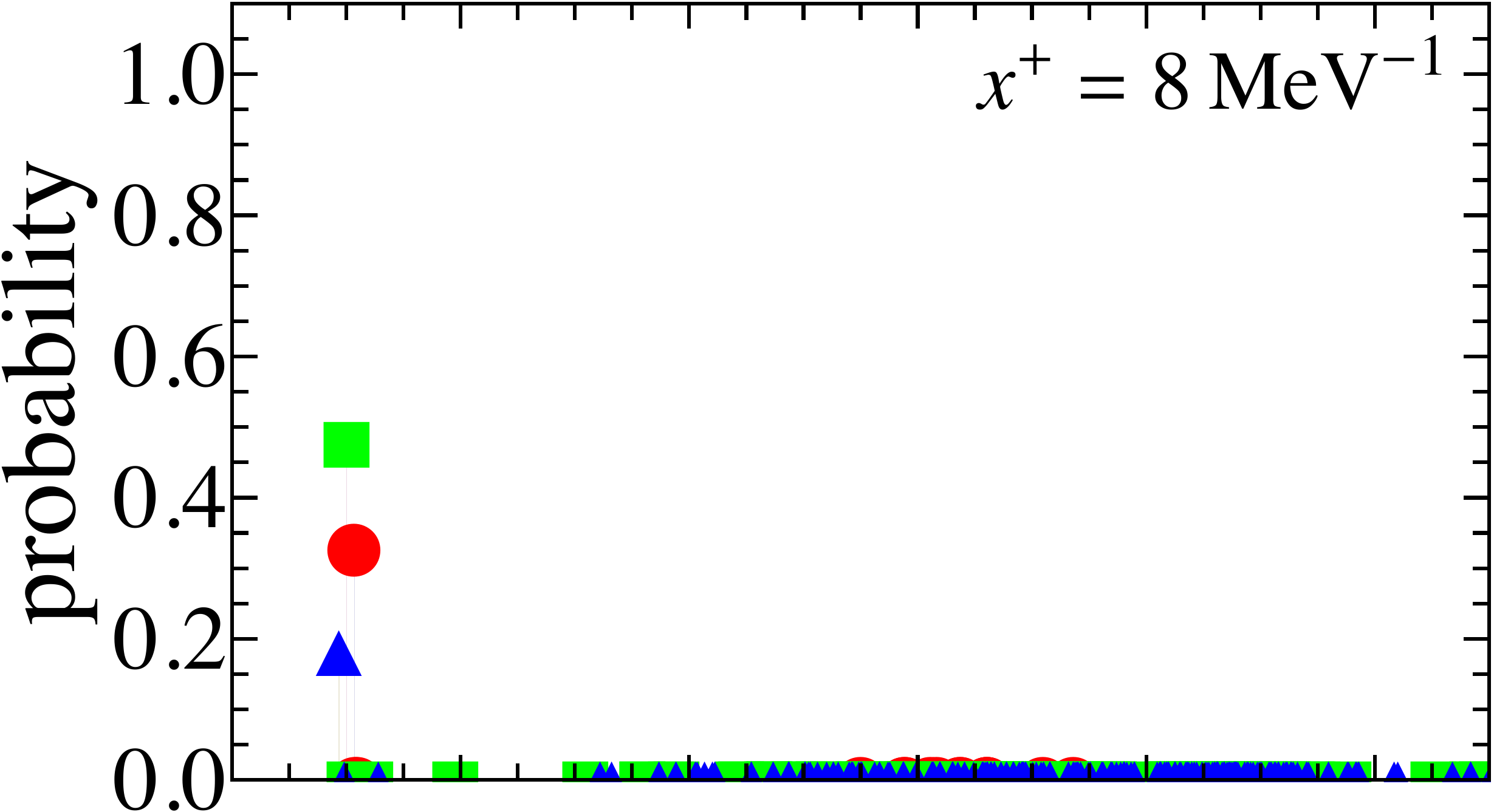}
\includegraphics[width=0.49\textwidth]{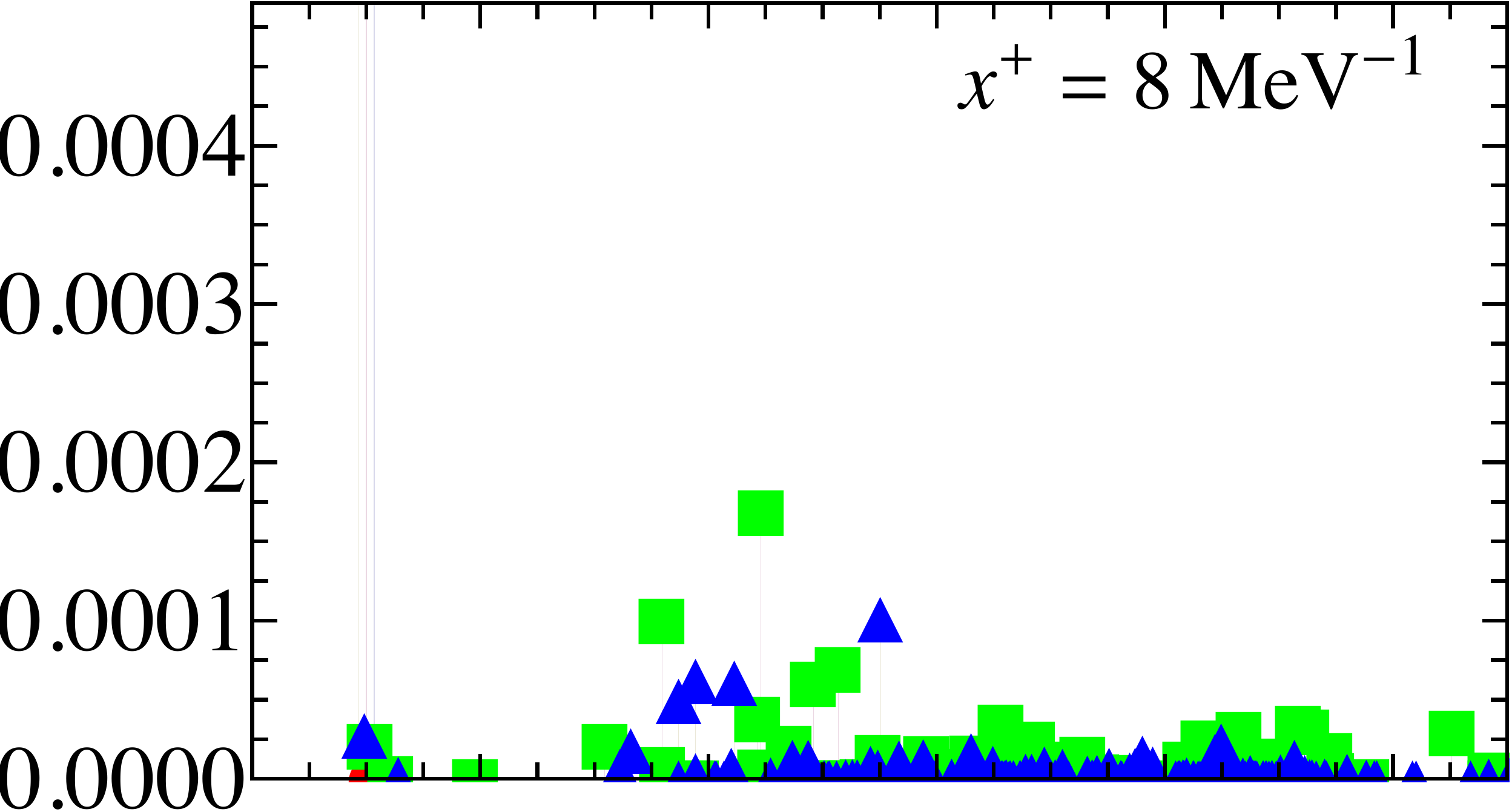}
\includegraphics[width=0.49\textwidth]{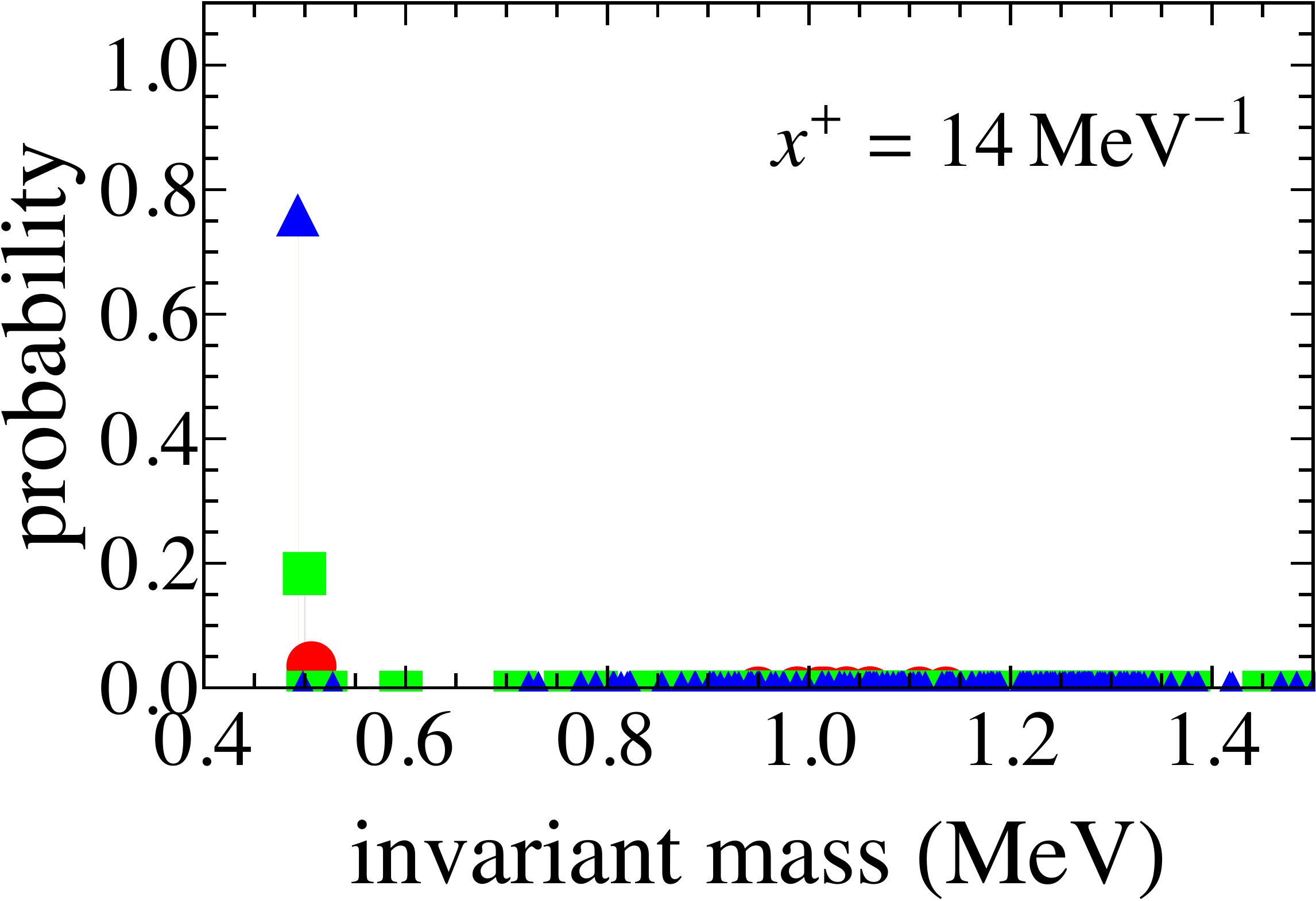}
\includegraphics[width=0.49\textwidth]{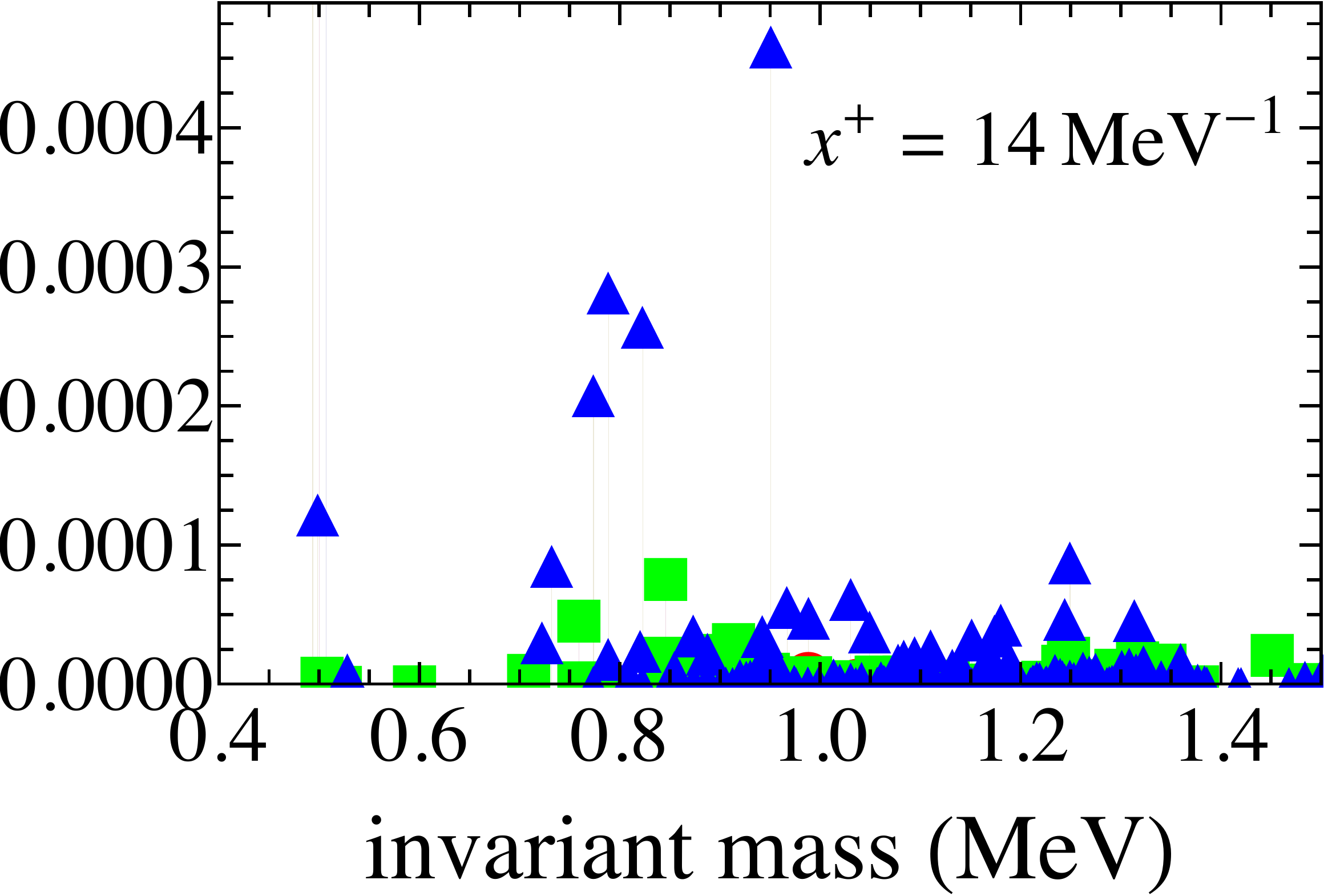}
\caption{(Color online) Non-perturbative time evolution of the single electron system in a laser field switched on at $x^+$= 0 using tBLFQ. 
From top to bottom, the panels in each row successively correspond to lightfront-time 
$x^+$= 4.0, 8.0, 14.0 MeV$^{-1}$.
Each dot represents an eigenstate of $P^-_{QED}$. The y-axis is the probability of finding this state, and the x-axis is the state's invariant mass. The left hand panels show the evolution of the three ground states 
in $K$=1.5, 3.5, 5.5 segments 
respectively and the right panels (with the y-axis expanded) show details of the excited state evolutions. The laser parameters used are: $a_0$=0.5 and $\omega=1.0$\,MeV. The electromagnetic coupling constant $\alpha{=}e^2/(4\pi)$ is 1/137.}
\label{fig:state_evol_np}
\end{figure*}

We first use BLFQ to solve for a set of QED mass eigenstates sufficient to include those that will be excited by the laser for its given strength and time duration; this set includes, for example, states with a range of longitudinal motions spanning several ``$K$-segments''\footnote{A $K$-segment denotes a group of BLFQ basis states with a common total longitudinal momentum $K$, see Ref.~\cite{Zhao:2013cma} for more details.}, since the laser injects longitudinal momentum into the system.
We then shine the laser (taken here as the classical source 
$e\mathcal{A}^-(x^-)=2 m a_0 \cos{(\omega x^-)}$ with $m=0.511$\,MeV, $\omega=1.0$\,MeV,
chosen so that it makes transitions between states 
with longitudinal momenta differing by 
$\Delta K$=2)
on the target (lowest mass eigenstate in lowest $K$-segment of
longitudinal motion) and time-evolve
the states in the basis of the original mass eigenstates according to Eq.~\ref{i_evolve}.
The basis uses $\sqrt{M_0 \Omega}$ = 0.511 MeV along with three $K$-segments 
with $K$=1.5, 3.5, and 5.5 
which allows the system to accelerate without (or with) excitation.  

Note that our chosen laser profile does not obey Maxwell's equations in vacuum. This is not an issue for us since we are interested here not in phenomenology but in a first demonstration of the framework of tBLFQ. Whether the background obeys Maxwell or not has no impact on our methods. With future developments of our formalism in mind, we note that a simple background field model obeying Maxwell would be a plane wave. However, it is also common to consider time-dependent electric fields, which do not obey Maxwell, as models of the focus of counter-propagating pulses~\cite{Dunne:2008kc}. Insisting on background field profiles which are both realistic (finite energy, pulsed in all four dimensions) and obey Maxwell's equations is a challenge, as very few such solutions exist in closed form. An exception is given in~\cite{IVAN}, and while there is nothing to stop us including such backgrounds in principle, doing so goes somewhat beyond the initial ``proof-of-concept" presented here.

With the chosen basis space and a laser field with $a_0$=0.5 we present results for the time evolution of the electron system in Fig.~\ref{fig:state_evol_np}. The tBLFQ evolution of the system is both coherent and nonperturbative - yielding the invariant mass distribution for electron + photon 
final states up to the (arbitrary) time the laser is switched off. Note that by the time at 
$x^+$=14 MeV$^{-1}$,
the initial 
$K$=1.5
ground state has nearly vanished. Predominantly, the electron has been accelerated to $K$=5.5.  States with larger invariant masses have acquired significant amplitudes especially in the mass range up to about 1 MeV.
Let us discuss the time evolution results in further detail.

In the initial system the only populated basis state is the single electron (ground) state in the $K{=}1.5$ segment. As time evolves, the background causes transitions from the ground state to states in the $K{=}3.5$ segment. Both the single electron states and electron-photon states are populated; the former represent the acceleration of the electron by the background, while the later represent the process of radiation. At times $x^+{=}4$~MeV$^{-1}$, the single electron state in $K{=}3.5$ becomes populated while the probability for finding the initial state begins to drop, as shown in the top left panel of Fig.~\ref{fig:state_evol_np}. In the top right panel, the populated electron-photon states begin forming a peak structure. The location of the peak is around the invariant mass of 0.8~MeV, roughly consistent with the expected value of $M_{\rm pk1}{=}\sqrt{P^-(K_i+k_{\rm las})}{=}0.78$~MeV.

Once the basis states in $K{=}3.5$ become populated, ``second" transitions to the $K{=}5.5$ segment become possible. This can be seen in the second row of Fig.~\ref{fig:state_evol_np}, at $x^+{=}8$~MeV$^{-1}$. In the left hand panel, one sees that the probability of the electron to remain in its ground state ($K{=}1.5$) is further decreased, the probability of it being accelerated (to $K{=}3.5$) is increased, and that the $K{=}5.5$ single electron state becomes populated. In the right hand panel, the electron-photon states in the $K{=}5.5$ segment also become populated as a result of the second transitions. A second peak arises here at the invariant mass of around $M_{\rm pk2}{=}\sqrt{P^-(K_i+2k_{\rm las})}{\sim}1.0$~MeV (distinct from at that ${\sim}0.8$~MeV, above, formed by the $K{=}3.5$ electron-photon states from the first transitions). The peak in the $K{=}5.5$ segment is at a larger invariant mass than that in the $K{=}3.5$ segment simply because the basis states in the $K{=}5.5$ segment follow from the initial state being excited {\it twice} by the background field, and thus receive more energy than states in the $K{=}3.5$ segment.

As time evolves further, the probability of finding a $K{=}5.5$ single electron exceeds that of finding a $K{=}3.5$ segment electron, see the bottom left panels in Fig.~\ref{fig:state_evol_np}. At this time, $x^+{=}14$~MeV$^{-1}$, the system is most likely be found in the $K{=}5.5$ single electron state, with probability $\sim$0.75. The probability for finding the $K{=}3.5$ single electron state is around~$0.2$ and the initial $K{=}1.5$ electron state is almost completely depleted. In the right hand panel, we see that the probability for finding $K$=5.5 electron-photon states increases with time. One also notices that at later times, the probability for $K$=3.5 electron-photon states also begin to drop. This is because (like the $K$=3.5 single electron) the $K{=}3.5$ electron-photon states are coupled to the $K{=}5.5$ single electron; as the probability of the $K$=5.5 single electron state increases, it ``absorbs" both the single electron and the electron-photon states in the $K$=3.5 segment. At $x^+$=14MeV$^{-1}$ we terminate the evolution process, as the system is already dominated by the single electron state in the maximum $K$-segment. Further evolution without artifacts would require bases with segments of $K{=}\{7.5, 9.5\ldots\}$.

\begin{figure}[!t]
\centering
\includegraphics[width=0.47\textwidth]{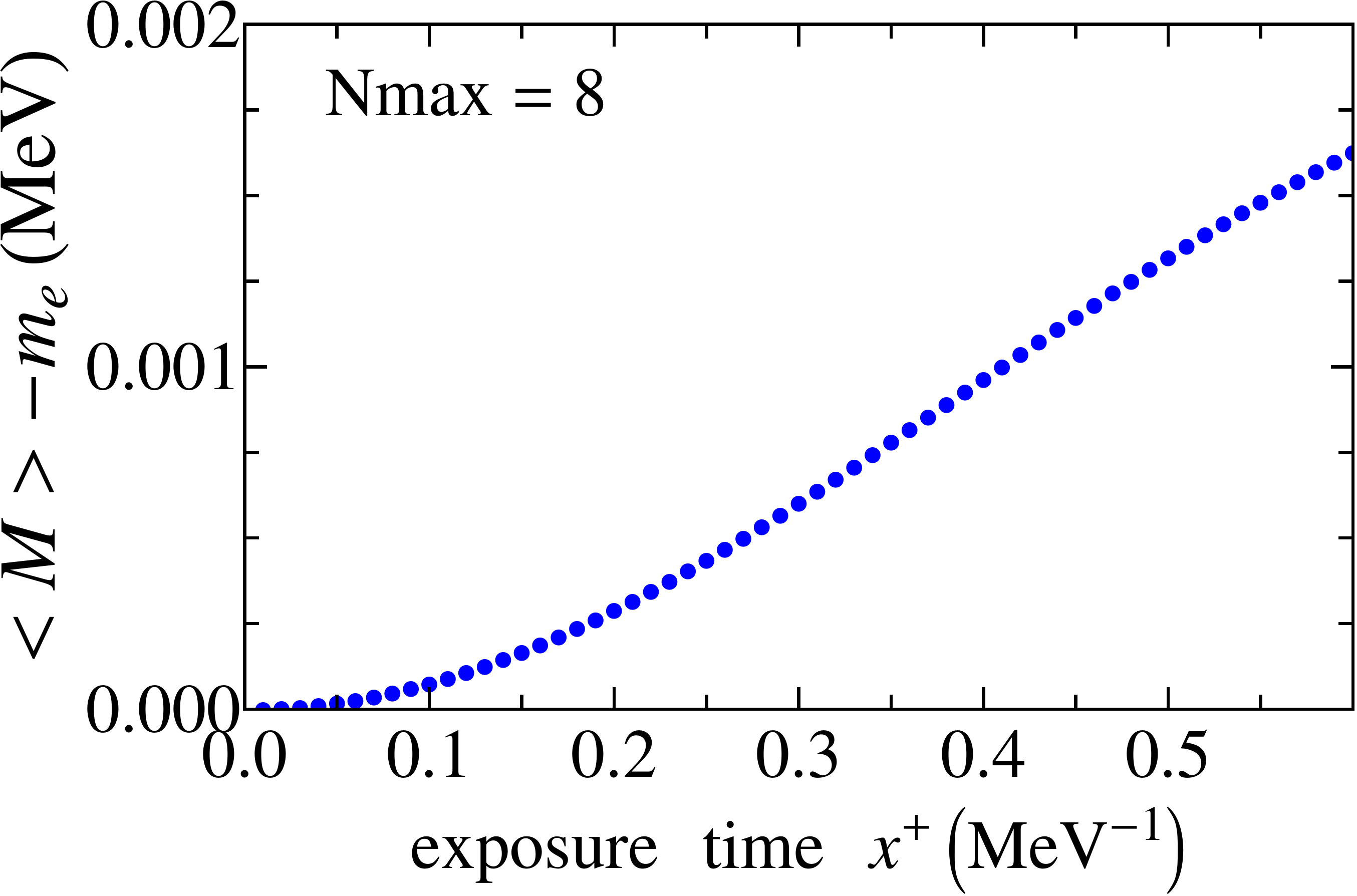}
\caption{Time evolution of the average invariant mass of the electron system calculated in tBLFQ basis space with $N_{\rm{max}}=8$. The Y-axis is the difference between the average invariant mass $\langle M\rangle$ of the system at $x^+$ and that of a single electron $m_e$. The X-axis is the (light-front) exposure time $x^+$. The laser parameters used are: $a_0$=10 and $\omega=1.0$\,MeV. The electromagnetic coupling constant $\alpha{=}e^2/(4\pi)$ is 1/137.}
\label{fig:state_evol_invmass}
\end{figure}

Since the states excited by the laser field encode all the information of the system, they can be employed to construct other observables. As an example, in Fig.~\ref{fig:state_evol_invmass} we present the evolution of the average invariant mass $\langle M\rangle$ of the system exposed in a laser field with $a_0=10$. The increase of the invariant mass with time reflects the fact that energy is pumped into the electron-photon system by the laser field.  This invariant mass can be accessed experimentally by measuring the momenta of both the final electron, $p^\mu_e$, and photon, $p^\mu_\gamma$ in a non-linear Compton scattering experiment. The invariant mass can be compared with the expectation value of $(p^\mu_e+p^\mu_\gamma)^2$ measured over many repetitions of the non-linear Compton scattering experiment.

These examples demonstrate: a) BLFQ is a suitable framework for evaluating non-perturbative stationary state solutions in quantum field theory; b) tBLFQ is able to generate first-principles-based, non-perturbative results for time-dependent strong processes in quantum field theory; and, c) the full quantum configuration (wavefunction) of the system is available for analyzing dynamical processes at any intermediate time.

Future developments will be made in several directions. The first direction is to extend the application range of BLFQ. Specifically, we are working on applying the non-perturbative light-front amplitudes to predict experimental observables such as Generalized Parton Distributions (GPDs). An example using the electron's light-front amplitudes is underway and will be reported elsewhere \cite{Chakrabarti:2014}.

The second direction is to make further improvement of tBLFQ itself. The initial step is to implement renormalization so that the BLFQ representation of the physical eigenspectrum of QED can be improved (and then used in tBLFQ calculations). Currently we are working on implementing a sector-dependent renormalization scheme within the BLFQ framework. The inclusion of higher Fock sectors in our calculation is also important, as it will not only result in more realistic representations of quantum states but will also allow for the description of a larger variety of processes, for example, multi-photon emissions. 

The third direction to be pursued is the extension of tBLFQ's range of applications. In the field of intense laser physics, the inclusion of transverse ($x^\perp$), longitudinal ($x^-$) and time ($x^+$) dependent structures to the background field will be used to more realistically model the focussed beams of next-generation laser facilities~\cite{IVAN}.  In addition to intense laser physics, we will also apply tBLFQ to relativistic heavy-ion physics, specifically the study of particle production in the strong (color)-electromagnetic fields of two colliding nuclei. Ultimately, the goal is to use tBLFQ to address strong scattering problems with hadrons in the initial and/or final states. As supercomputing technology continues to evolve, we envision that tBLFQ will become a powerful tool for exploring QCD dynamics.

We acknowledge fruitful discussions with Kirill Tuchin, Paul Wiecki, Yang Li and Guy de Teramond.
This work  was supported in part by a DOE Grant Nos. DE-FG02-87ER40371, DESC0008485 (SciDAC-3/NUCLEI), DE-FG02-93ER40771, and by DOE Contract No. DE-AC02-76SF00515 and by US NSF grant 0904782. A.~I.\ is supported by the Swedish Research Council, contract 2011-4221. Computational resources were provided by the National Energy Research Supercomputer Center (NERSC), which is supported by the Office of Science of the U.S. Department of Energy under Contract No. DEAC02-05CH11231. This paper is SLAC-PUB-15903.





\end{document}